\documentclass[aps,pra,amsmath,amssymb,reprint,letterpaper,raggedbottom]{revtex4-1}

\usepackage{graphicx}
\usepackage{color}
\usepackage{natbib}

\begin{document}

\newcommand*{\de}{$\Delta\varepsilon\:\:$}
\title{Utilizing a Perturbative Inverse of the Hamiltonian to Improve Lanczos Diagonalization}

\author{Jeffrey R. Keyes}
\affiliation{Department of Chemistry, Wesleyan University, Middletown,
  CT 06459}

\author{Carlos A. Jim\'enez-Hoyos}
\affiliation{Department of Chemistry, Wesleyan University, Middletown,
  CT 06459}

\begin{abstract}
The use of the shifted inverse Hamiltonian in the Lanczos algorithm is examined.
We present a perturbative inverse and demonstrate that the Krylov subspace
formed from it is better suited to approximating the ground state wavefunction
compared with the Hamiltonian itself. We further demonstrate that the
perturbative inverse provides an adequately accurate approximation to the
inverse Hamiltonian at second order. Naturally, the quality of the results
depend on the quality of the initial guess, here we use the Hartree-Fock, CISD,
MP-PT1, and EN-PT1 wavefunctions. Lastly, we examine the scaling of the method
where we find that accurate results can be found for similar cost as PT5. 

\end{abstract}

\maketitle

\section{Introduction}
It is very common in theoretical physics that we seek the eigenvector of some
matrix corresponding to the smallest eigenvalue or sometimes just the eigenvalue
itself. When the dimension of the underlying vector space is too large to store
the matrix, which is common when diagonalizing the Hamiltonian where the
dimension of the full configuration interaction(FCI) space scales exponentially
with the number of particles, it is standard practice to turn to the Lanczos
algorithm\cite{lanczos_iterative_1950,chen_systematic_2022,zhuravlev_cumulant_2022}
or Davidson
algorithm\cite{davidson_iterative_1975,liu_simultaneous_1978,windom_iterative_2023}
where we try to build a subspace of the Hilbert space that contains a good
approximation of the vector we seek.

Both of these methods have drawbacks, the Lanczos algorithm has the advantage
that all elements required can be gathered without storing any FCI vectors but
the algorithm produces vectors that are nearly orthogonal to the exact
wavefunction and will formally diverge in a complete basis. Davidson does not
have this problem, but only works well if the matrix is in a diagonally dominant
representation or pre-conditioned well. In this paper, we intend to improve on
the Lanczos method by building a Krylov subspace that is better suited to
approximating the ground state of the Hamiltonian.

The reason that the Lanczos vectors become nearly orthogonal to the exact
wavefunction is because the Krylov subspace tends towards the eigenvector with
the largest eigenvalue, sampling the upper end of the spectrum more than the
lower end. This is also what causes the worsening performance with a more
complete orbital basis, since the spectrum of the Hamiltonian is not bounded
from above, a larger orbital basis will be able to include higher energy states
that will allow the Lanczos vectors to move further away from the ground state.
In fact, if the orbital basis is complete, then the norm of the Lanczos vectors
would diverge. In this paper, we circumvent this problem by employing a shifted
inverse Hamiltonian. This will cause the Lanczos vectors to tend towards the
ground states wavefunction and produce a Krylov subspace that can produce better
approximations of the desired state. Importantly, this method will not fomally
diverge in a complete orbital completeness. To make the use of the inverse
Hamiltonian computationally feasible, we introduce a perturbative inverse that
can approximate the exact inverse.

The inverse of the Hamiltonian has been explored in the past,
Nakatsuji\cite{nakatsuji_inverse_2002} looked at the inverse Schr\"odinger
equation and suggested augmenting the Krylov subspace for the Hamiltonian with
that from the inverse. His study was limited to the case of the hydrogen atom,
but demonstrated that the approach can work. Our work is distinguished in a few
important ways, in his study of the hydrogen atom, he used the kinetic energy as
the perturbation which clearly is not well suited for the molecular case, we
instead test both the M\o ller-Plesset and Epstein-Nesbet partitioning. We also
discuss more precisely why the Krylov subspace generated by the inverse
Hamiltonian is better than the Hamiltonian. Nakatsuji explained the choice by
noting "since the inverse Schr\"odinger equation is equivalent to the
Schr\"odinger equation, the inverse Krylov sequence should also be important in
the expansion of the exact wave function", but made no comparison of these two
Krylov subspaces. Lastly, we explore the effect of using different shift
parameters(see below) on the quality of the resulting Krylov subspace.

The perturbative inverse has also been considered more recently by Windom and
Bartlett\cite{windom_iterative_2023} where they explore the relationship between
the Brillouin-Wigner PT resolvent operator and the preconditioner used in
diagonalization methods in order to understand and improve the convergence
behavior of the preconditioned Lanczos and Davidson procedures.

The paper is organized as follows, in Sec. \ref{sec:Theory}, we review the
Lanczos method and discuss how the inverse Hamiltonian can be useful. We also
present the perturbative inverse that we used and discuss the computational
scaling of the method. In Sec. \ref{sec:Results}, we demonstrate the
capabilities and features of the method primarily through calculations on LiH.
In Sec. \ref{sec:Conclusion}, we provide some concluding thoughts and discuss
possible future directions of this project.

\section{Theory}\label{sec:Theory}
\subsection{Power Method and Lanczos Method}
The Lanczos method is a fairly simple extension of the power method, which we
review here for completeness. Suppose we have a matrix, $A$, from which we seek
the eigenvector that corresponds to the largest eigenvalue. We can generate this
vector by observing the following relationship:
\begin{equation}
  A^n|x\rangle = \sum_i a_i^n |a_i\rangle\langle a_i|x\rangle
\end{equation}
where $|x\rangle$ is some vector and $A|a_i\rangle = a_i|a_i\rangle$. We can see
here that as we repeatedly apply the matrix onto the vector, the contribution of
the eigenvector with the largest magnitude eigenvalue will dominate so long as
$\langle a_i|x\rangle\neq 0$. We can then write the eigenvector as the following
limit:
\begin{equation}
  |a_{max}\rangle = \lim_{n\to\infty} \frac{A^n|x\rangle}{\langle x|A^{2n}|x\rangle}
  \label{eq:amax}
\end{equation}
where we introduce the denominator to ensure that the norm does not diverge.

While this relationship can acquire the eigenvector in question, taking the
above limit may require too many steps. We can improve the convergence by
utilizing the various terms in the limit as a basis in which we can diagonalize
the matrix. For this, we define the Krylov subspace $\mathcal{K}_n =
\text{span}\{A^m|x\rangle | m = 0,1,2,\cdots,n-1 \}$. The vector that we seek
will be the eigenvector of $A$ in the $\mathcal{K}_n$ subspace with the largest
eigenvalue as $n$ goes to infinity. This is the most basic form of the Lanczos
method. One of the major advantages of this algorithm is that the elements of
$A$ in the subspace can be constructed without ever carrying the matrix $A$ or
any vectors of the dimension of $A$. The only requirement is the ability to
multiply the operator onto some vector to calculate $\langle
x|A^{2m+1}|x\rangle$. It should be noted that the power method often makes many
vectors that are nearly linearly dependent, so this procedure usually involves a
scheme to orthogonalize the vectors to avoid numerical difficulties.

\subsection{Inverse Power Method}
While the power method works well for gathering the eigenvector with the largest
eigenvalue, we often seek the one with the smallest eigenvalue. If we tried
using the Krylov subspace from the power method to find this vector, each
subsequent vector moves further away from the vector that we seek, so more steps
may have to be taken before a subspace with the vector of interest is found
inside. We can circumvent this problem by building the Krylov subspace out of
the inverse operator by noting:
\begin{equation}
  (A-\varepsilon I)^{-n}|x\rangle = \sum_i \frac{1}{(a_i-\varepsilon)^n} |a_i\rangle\langle a_i|x\rangle
\end{equation}
where we introduce a real valued shift, $\varepsilon$, to the operator and $I$
is the identity. If we apply this relationship the same as we did with Eq.
\ref{eq:amax}, we can get the eigenvector with eigenvalue closest to
$\varepsilon$. Similarly, we can build the Krylov subspace, $\mathcal{K}_n =
\text{span}\{(A-\varepsilon I)^{-m}|x\rangle | m = 0,1,2,\cdots,n-1 \}$, where
we can diagonalize the operator, $A$, to find the eigenvector we seek. Using the
inverse operator should accelerate convergence to the vector we want.

\subsection{The Problem at Hand}
We wish to apply the above concepts to finding solutions of the molecular
Schr\"odinger equation:
\begin{equation}
  H_N|\Psi\rangle = E_{corr}|\Psi\rangle
\end{equation}
where $H_N$ is the normal ordered Hamiltonian and $E_{corr}$ is the correlation
energy. In the majority of situations, researchers are interested in the ground
state and perhaps the first few excited states. For this we can build the Krylov
subspace $\mathcal{K}_n = \text{span}\{(H_N-\varepsilon I)^{-m}|0\rangle | m =
0,1,\cdots,n-1\}$ where $\varepsilon$ is a parameter that will be set to a value
below the exact correlation energy and $|0\rangle$ is any approximate
wavefunction that is not orthogonal to the ground state. We can then build the
Hamiltonian in this subspace to approximate the ground state. In this paper, we
will keep $\varepsilon$ the same for each iteration but there is no real
requirement of this. It may be advantageous to adjust the shift parameter at
each step, perhaps using information about the energy at each step to make a
better guess. If an excited state is sought, then an initial wavefunction that
is orthogonal to the ground state can be used. If it has a different symmetry,
this can be trivial, if not it may be difficult finding an appropriate state. In
this paper, we will focus on finding the ground state and leave excited states
to future investigations.

An important fact to point out is that the spectrum of the Hamiltonian is not
bounded from above. This means that when building the Krylov subspace, the
vectors would diverge in a complete orbital basis. For practical calculations,
this means that the Lanczos algorithm will perform worse when using larger basis
sets. However, when using the inverse Hamiltonian, the Krylov subspace is always
well defined so long as the shift parameter in not in the spectrum of the
Hamiltonian even when using a complete orbital basis. Further, from the HVZ
theorem\cite{simon_geometric_1977}, we know the spectrum of the molecular
Hamiltonian is bounded from below, so if the shift parameter is chosen to be
below the exact energy, then the ground state is guaranteed to be the closest
state.

Obviously, in order to utilize this method, we need to be able to invert the
Hamiltonian which in general is an infinite body operator. The simplest way to
the invert the Hamiltonian is to build the matrix in the FCI space and invert
directly. However, due to the exponentially increasing size of the Hilbert space
with number of electrons, this would be impossible in most circumstances.
Instead, from the identity $(A+B)^{-1} = A^{-1} - A^{-1}B(A+B)^{-1}$, we get the
following relationship:
\begin{multline}
  \label{eq:inv}
  (H_N - \varepsilon I)^{-1} = \sum_{n=1}^\infty (-1)^{n-1}(H_0-\varepsilon I)^{-1}\\
  \times[V(H_0-\varepsilon I)^{-1}]^{n-1}
\end{multline}
where $H_N = H_0 + V$ and $H_0 - \varepsilon I$ is trivially invertible. This
relationship defines a perturbative inverse where terminating the series at a
finite order gives us an approximate inverse of the Hamiltonian. Importantly,
diagonalizing the Hamiltonian in this basis requires calculating the terms
$\langle (H_N - \varepsilon I)^{-m}H_N(H_N - \varepsilon I)^{-m}\rangle$, which
can be done diagrammatically without ever building the operator in the FCI basis
or storing the Lanczos vectors.\cite{shavitt_many-body_2009} It should be noted
that since Lanczos vectors are not extensive, then this method as described is
not extensive.

A simple modification that can be done to regain extensivity is to use the
Lanczos vectors as approximations of the ground state wavefunction and evaluate
the energy projectively.
\begin{equation}
  \label{eq:proj}
  E_n = \frac{\langle 0|H_N(H_N-\varepsilon I)^{-n}|0\rangle}
  {\langle 0|(H_N-\varepsilon I)^{-n}|0\rangle}
\end{equation}
Since all unlinked terms are removed in the projection onto the reference state,
this is extensive and since the Lanczos vectors approach the exact ground state
with increasing $n$, we have $E_{corr} = \lim_{n\to\infty}E_n$. Clearly, this
has some relation to perturbation wavefunctions and greens function solutions
when the shift parameter is the HF energy, but we will not explore that here.

In principle, any function of the inverse could be used to build the Krylov
subspace, so we will also test the Krylov space generated by the operator,
$\exp[(H_N-\varepsilon I)^{-1}]$. The benefit of this choice is that it will
include contributions from higher powers of the inverse Hamiltonian and the cost
of using each subsequent vector does not increase since 

\begin{equation}
  (\exp[(H_N-\varepsilon I)^{-1}])^n = \exp[n(H_N-\varepsilon I)^{-1}]
\end{equation}
So if one can calculate the element $\langle H_N \exp[n(H_N-\varepsilon
I)^{-1}]\rangle$, then the calculation can be done using any of the vectors from
the start. Since these vectors get closer to the exact solution as $n$
increases, one can skip to using a vector with large $n$ without constructing
the ones before. Currently, we are not aware of any way to calculate these
values without expanding the exponential.

\subsection{Scaling}
For this method to be useful, it must have a reasonable computational scaling.
The most expensive step of this process is the calculation of the last
Hamiltonian element. Specifically, if we truncate the perturbative inverse at
order $n$ and use up to powers of $m$ to build the Krylov subspace, the most
expensive calculation is the element:
\begin{multline}
  \langle 0|[(H_0-\varepsilon)^{-1}[V(H_0-\varepsilon)^{-1}]^{n-1}]^m\\
  \times V[(H_0-\varepsilon)^{-1}[V(H_0-\varepsilon)^{-1}]^{n-1}]^m |0\rangle
\end{multline}
which has a cost similar to PT$(2nm+1)$. As we will discuss below, letting $n=2$
and $m=1$ or $m=2$ produces good results, so the cost would be similar to PT5
and PT9 respectively. This calculation can be done using diagrammatic techniques
similar to regular perturbation theory with the exception that here we must
include the unlinked diagrams as well. 

In principle, the cost could be reduced by projecting each Lanczos vector into
some subspace of the Hilbert space and carrying the amplitudes. This would make
going to higher $m$ more feasible, however this will limit the accuracy of the
method since it will only produce vectors in that limited subspace. For example,
if we project the vectors in the basis of singles and doubles, then the method
will approximate the CISD energy rather than the FCI energy.

The alternative method involving projections has a lower scaling when evaluating
the energy, which is reduced to that of PT$(nm+1)$, so more steps can be taken.
However, since this is equivalent to the power method rather that the Lanczos
method, the convergence will be slower.

\subsection{Implementation}
This method has been implemented using in-house code building all terms in the
FCI basis. This will limit the size of systems that we can currently perform
calculations on. Integrals and initial wavefunctions were calculated using the
PySCF software package. The solutions to the generalized eigenvalue problem were
found using the scipy linear algebra python package. All calculations were
performed on LiH using the experimental bond length of 1.5949\r{A} with an
STO-3G basis unless otherwise specified.

\section{Results and Discussion}\label{sec:Results}

\begin{figure}[tb]
  \includegraphics*[width=\columnwidth]{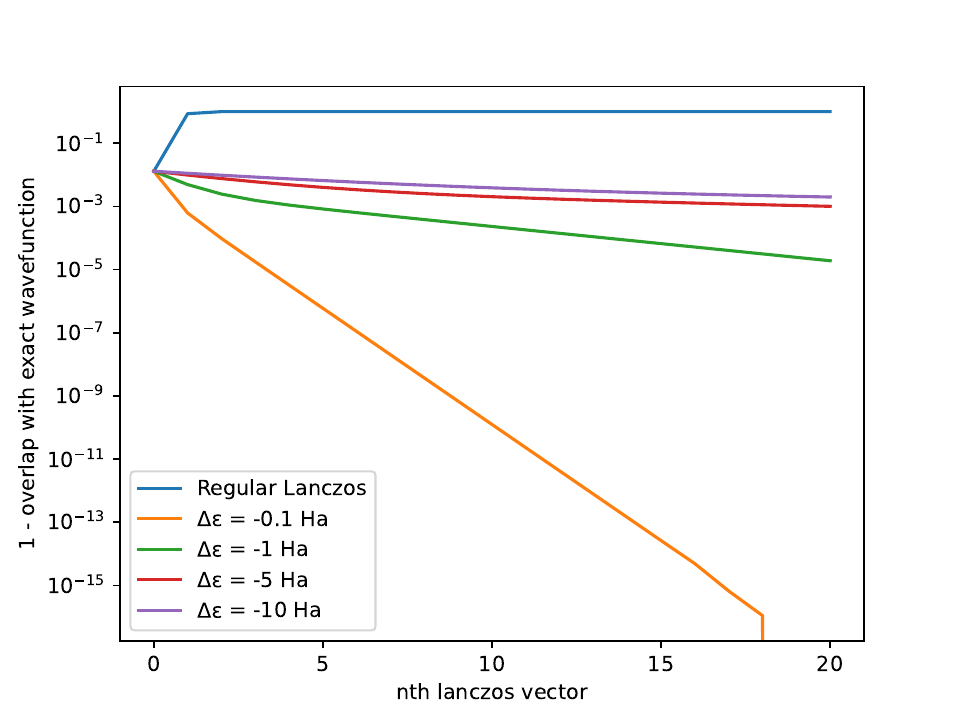}
  \caption{Comparing the overlap of the nth Lanczos vector and the exact
  wavefunction using the Hamitonian and inverse Hamitonian with various \de.}
  \label{fig:inv_comp}
\end{figure}

We begin by demonstrating the assertion made in the previous section that using
the inverse Hamiltonian yields Lanczos vectors that are much closer to the exact
wavefunction. A value that will be used a lot in this paper is the difference
between the ground state correlation energy and the shift parameter, so we
define \de $ = \varepsilon - E_{corr}$. In Fig. \ref{fig:inv_comp}, we plot  
the norm of difference between the first twenty lanczos vectors and the FCI
wavefcuntion. The zeroth order wavefunction used here is the HF wavefunction
with a small random vector added to it. Here, the Lanczos algorithm using the
Hamiltonian quickly produces vectors that are nearly orthogonal to the FCI
wavefunction with the overlap dropping to $10^{-8}$ after only 3 steps. In
contrast, performing the Lanczos algorithm with the inverse Hamiltonian causes
the vectors to quickly approach the FCI wavefunction. Even when \de is set to
-10 Ha, it requires only 4 steps to have an overlap of 0.95. It's also clear
that as the \de approaches zero that the convergence onto the FCI wavefunction
is accelerated. When using a \de of -0.1 Ha, the first Lanczos vector already
has an overlap of 0.9986. From this data, it is clear that the Krylov subspace
formed from the inverse Hamiltonian will be better suited to approximating the
FCI wavefunction.

\begin{figure}[tb]
  \includegraphics*[width=\columnwidth]{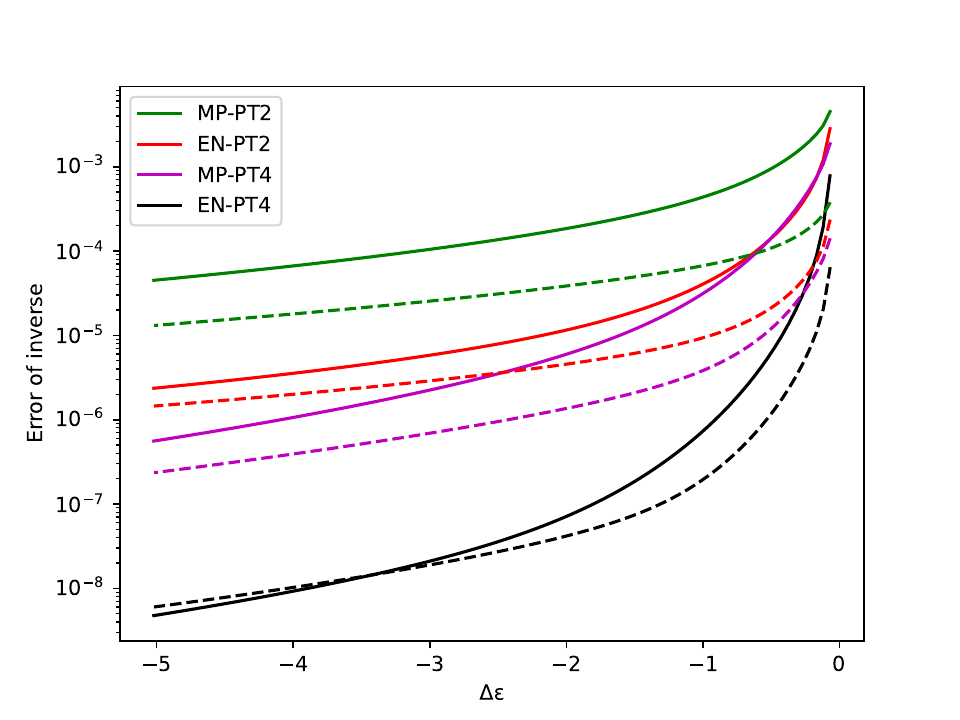}
  \caption{Plot of error in inverse Hamiltonian using 2nd and 4th order MP-PT
  and EN-PT inverses. Solid lines are using STO-3G, dotted lines are using
  6-31G. See text for definition of error.}
  \label{fig:err}
\end{figure}

While the inverse Hamiltonian produces a better subspace, we need to ensure that
this is still the case when using an approximate inverse. Using Eq.
\ref{eq:inv}, we can build approximate inverses by truncating at finite order,
here we look at the cases of $n=2$ and $n=4$ using both M\o ller-Plesset(MP-PT)
and Epstein-Nesbet(EN-PT) perturbation theories. In Fig. \ref{fig:err}, we show
the error of using perturbation theory compared to building the inverse using
the full matrix. For this, we define the error:
\begin{equation}
  Error = \frac{||I-H^{-1}H||}{N}
\end{equation}
where $||\cdot||$ is the Frobenius norm and $N$ is the dimension of the Hilbert
space. The denominator is introduced to make the error comparable between basis
sets. Here, the results are as one may expect, larger $n$ produces a more
accurate inverse and EN-PT tends to outperform MP-PT when truncated to the same
order. The other important observation here is that as the \de approaches zero,
all inverses tend to become less accurate. This is not surprising since as the
smallest eigenvalue of $H-\varepsilon$ approaches zero, the inverse becomes more
ill-conditioned, so Eq. \ref{eq:inv} will converge slower. It should be noted
that errors in the inverse do not compromise the utility of the method, an
approximate inverse will still produce vectors that are closer to the FCI
wavefunction, so it can still produce a useful Krylov subspace. 

\begin{figure}[tb]
  \includegraphics*[width=\columnwidth]{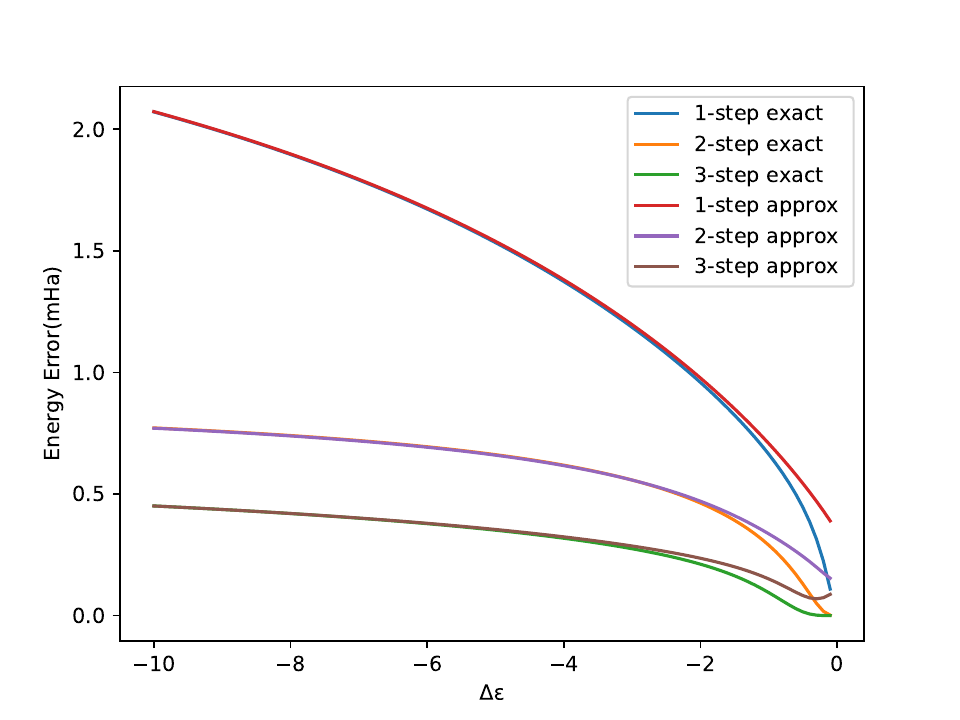}
  \caption{Comparing the error in ground state energy using the exact and
  perturbative inverse at various \de. Initial guess is the HF wavefunction and 
  using second order MP inverse.}
  \label{fig:exactvsPT}
\end{figure}

In order to evaluate the quality of the Krylov subspace formed with the
approximate inverse, we compare the Lanczos method using the exact and
approximate inverse. In Fig. \ref{fig:exactvsPT}, we show the error in the
Lanczos energy starting from the HF wavefunction for the first three steps for
various \de values where the approximate inverse was built to second order using
the MP partitioning. All calculations produced an energy within about 2 mHa of
the correct answer even when \de is 10 Ha. As is to be expected, including more
Lanczos vectors produces significant improvements to the energy and as \de
approaches zero, the calculation using the exact inverse improves monotonically.
The calculations using the approximate inverse matches the exact when \de is
large, but deviates as \de approaches zero as a consequence of inaccuracies in
the inverse as discussed above. We found in our testing that using the
approximate inverse often produces energies above the exact method, but this is
not always the case, occasionally the approximate method produces a lower
energy. Regardless, it is clear from this data that the perturbative inverse can
produce a Krylov subspace of similar quality to the exact inverse.

\begin{figure}[tb]
  \includegraphics*[width=\columnwidth]{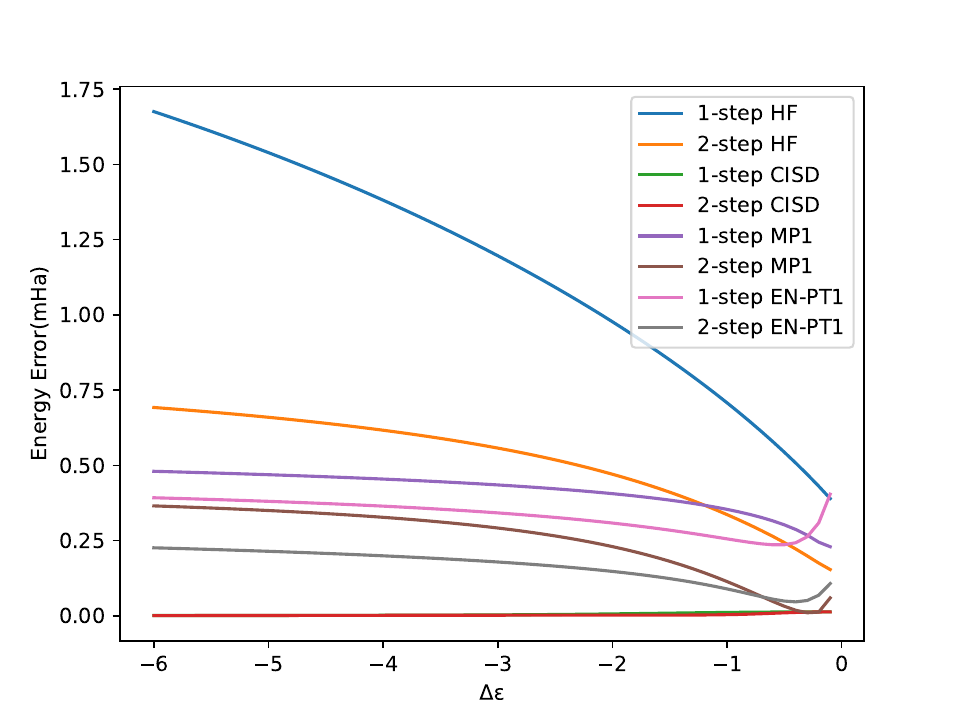}
  \caption{Comparing the error in ground state energy using different initial guesses at various \de.}
  \label{fig:guess}
\end{figure}

A major factor determining the quality of the result is the choice of initial
guess, here we compare four of them, the HF, CISD, MP-PT1, and EN-PT1
wavefunctions. In Fig. \ref{fig:guess}, we compare these four initial guesses
for LiH and the EN-PT1 calculation also used EN-PT to build the approximate
inverse. As to be expected, starting from a more accurate wavefunction will
improve the overall results, though this gap closes significantly as \de
approaches zero. It is also common that smaller \de is not always better since
inaccuracies in the inverse become worse as \de goes to zero. This suggests
that an optimization of the shift parameter may be the best way to use this
method.  When comparing the MP-PT calculations to EN-PT, we typically see
better performance with EN-PT for most \de, which is because EN-PT produces
more accurate wavefunctions and smaller errors in the inverse. Though these
plots cross for small \de, suggesting that the MP inverse is better when the
Hamiltonian is poorly conditioned.



\begin{figure}[tb]
  \includegraphics*[width=\columnwidth]{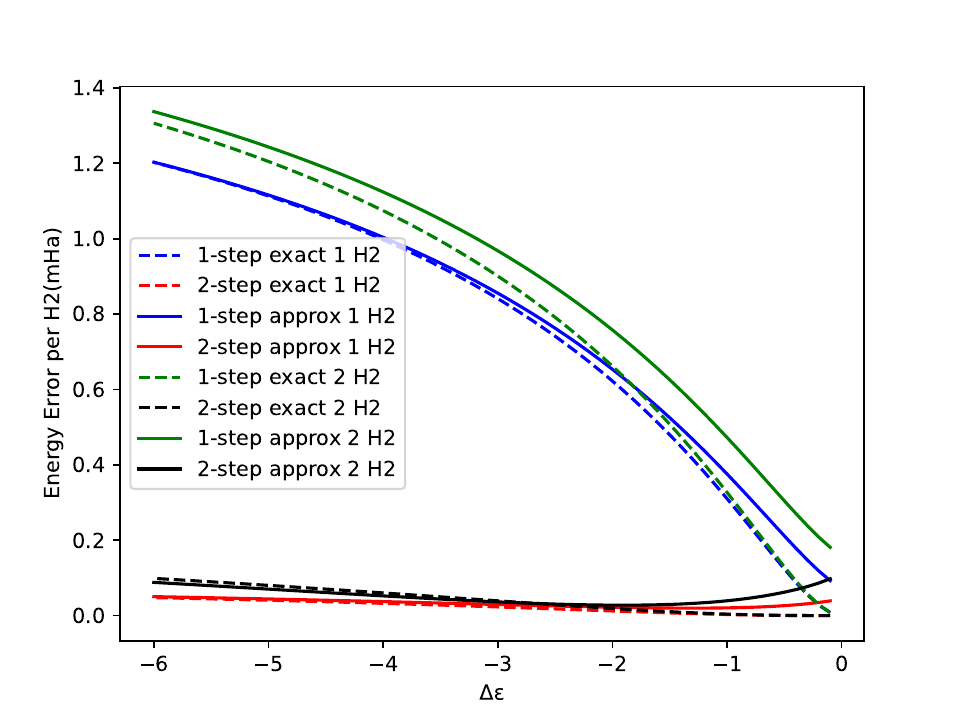}
  \caption{Comparing the error of ground state energy of H2 and 2H2. Initial guess is HF wavefunction.}
  \label{fig:exten}
\end{figure}

It is important when proposing a method that the extensivity is examined, in
Fig. \ref{fig:exten} we compare calculations done on H$_2$ and 2 H$_2$ molecules
separated by 100\r{A} using the s functions from the cc-pVDZ basis set. We see
here that there are two major effects of increasing the system size. First,
since the overlap of the HF and FCI wavefunctions decays as system size
increases and the lack of extensivity discussed above, then there will be a
larger error for all shift parameters. This can be seen by comparing the curves
using the exact inverse, there is an error for large \de, but when \de is small,
the two curves join since the method is formally exact when \de is zero. The
second error that we get is that the quality of the perturbative inverse
decreases for the larger system. This may be due to the presence of unlinked
terms in the perturbative inverse, but both curves recombine with the exact
inverse for large enough $\Delta\varepsilon$. Both of these errors are
significantly reduced when using the second Lanczos vector. A more detailed
study is needed to determine how significant this problem is.

\begin{figure}[tb]
  \includegraphics*[width=\columnwidth]{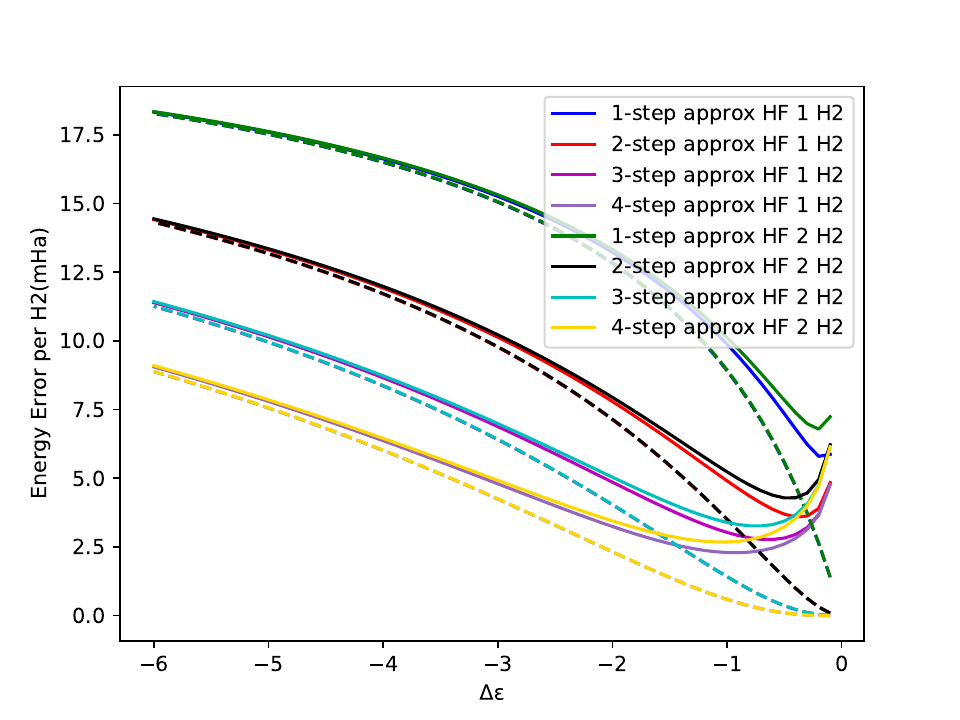}
  \caption{Comparing the error of ground state energy of H2 and 2H2. Initial
  guess is HF wavefunction. Solid lines use approximate inverse and dotted lines
  use the exact inverse.}
  \label{fig:exten_alt}
\end{figure}

As discussed in Sec. \ref{sec:Theory}, extensivity can be recovered by
projecting the lanczos vectors onto the reference state. In Fig.
\ref{fig:exten_alt}, we perform the same calculation but this time evaluating
the energy according to Eq. \ref{eq:proj}. The extensivity can be seen be the
fact that the dotted plots using the exact inverse are identical. Further, the
energy using the approximate inverse have a similar error for large \de, which
was not the case in Fig. \ref{fig:exten}, so the extensivity in the inverse has
been somewhat alleviated. There is still a larger error in the inverse for the
larger systems for small \de. Comparing the axes of Fig. \ref{fig:exten} and
\ref{fig:exten_alt}, it is clear that the projective method converges slower.
This is not surprising since the projective method uses only one Lanczos vector
as opposed to linear combinations of multiple Lanczos vectors. It remains to be
seen how these two approaches fair for large systems.

\begin{figure}[tb]
  \includegraphics*[width=\columnwidth]{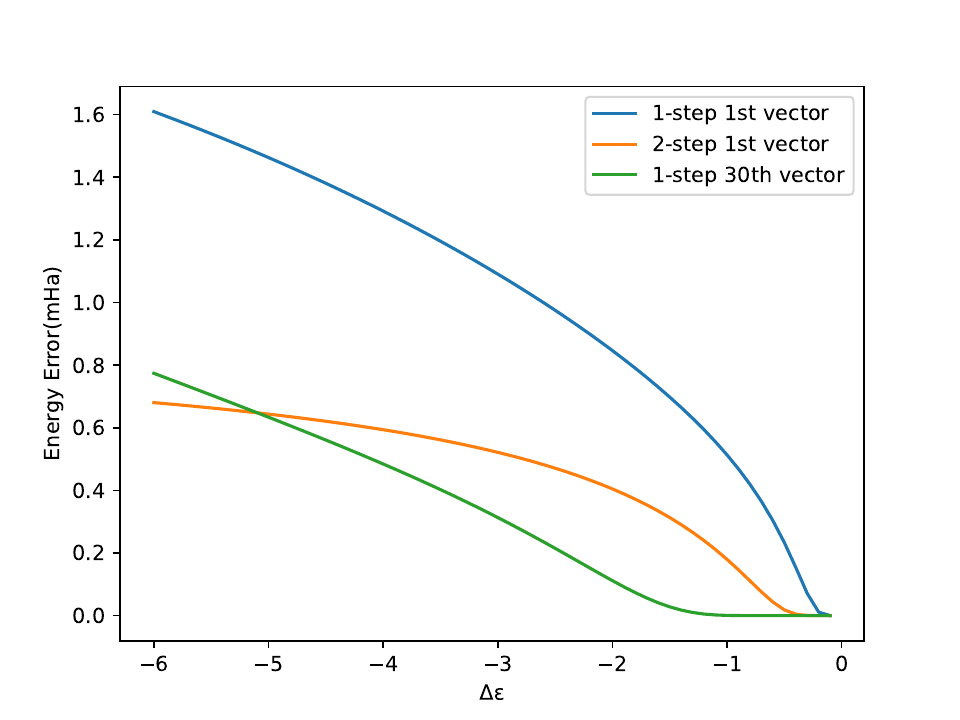}
  \caption{Comparing the error in ground state energy using the first and
  thirtieth Lanczos vectors of the exponential basis at various \de.}
  \label{fig:exp}
\end{figure}

Lastly, we wish to demonstrate the potential of using exponential basis
discussed above. The advantage of this basis is that since the cost of producing
any Lanczos vector is the same, the basis vectors can be produced in any order
without affecting computational cost. As an example, in Fig. \ref{fig:exp}, we
compare using the first and the 30th vector in LiH. Here, we see that using the
30th vector greatly improves the energy when compared to using the first two
Lanczos vectors. Unfortunately, we don't currently know of anyway to make
practical use of this knowledge.

\section{Conclusion}\label{sec:Conclusion} In this paper, we have proposed using
a perturbative inverse to build a Krylov subspace for the Lanczos algorithm. We
have demonstrated that this perturbative inverse can produce a Krylov subspace
of similar quality as the that formed from the exact inverse. Further, we have
shown that the method is capable of getting good approximations of the FCI
energy even without a good guess of the energy.

Future directions for this method include producing a more efficient
implementation to evaluate the capabilities of this method on larger systems.
Further, modifications that make finding excited states easier would be needed, this could be
accomplished by different starting guesses, like CIS wavefunctions or
$\Delta$SCF wavefunctions. Excited states could also be gathered by scanning the
shift parameter above the ground state energy, though the perturbative inverse
may make it difficult to identify when the excited states has be found.

\bibliography{Inverse_Lanczos.bib}
\bibliographystyle{plain}

\end{document}